\documentclass[floatfix,aps,prl,reprint,groupedaddress]{revtex4-1}
\usepackage{graphicx}
\usepackage{amsmath}

\begin{document}
\title{Magnetic Trapping of Cold Bromine Atoms}

\author{C. J. Rennick}
\author{J. Lam}
\author{W. G. Doherty}
\author{T. P. Softley}
\email{tim.softley@chem.ox.ac.uk}

\affiliation{Department of Chemistry, University of Oxford, Chemistry Research 
Laboratory, Mansfield Rd, Oxford, OX1 3TA}

\date{\today}

\begin{abstract}
Magnetic trapping of bromine atoms at temperatures in the milliKelvin regime is demonstrated for the first time.
The atoms are produced by photodissociation of Br$_2$ molecules in a molecular beam.
The lab-frame velocity of Br atoms is controlled by the wavelength and polarization of the photodissociation laser.
Careful selection of the wavelength results in one of the pair of atoms having sufficient velocity to exactly cancel that of the parent molecule, and it remains stationary in the lab frame.
A trap is formed at the null point between two opposing neodymium permanent magnets.
Dissociation of molecules at the field minimum results in the slowest fraction of photofragments remaining trapped.
After the ballistic escape of the fastest atoms, the trapped slow atoms are only lost by elastic collisions with the chamber background gas.
The measured loss rate is consistent with estimates of the total cross section for only those collisions transferring sufficient kinetic energy to overcome the trapping potential.
\end{abstract}


\maketitle

Existing cold and ultracold atom sources are primarily based on laser Doppler cooling of alkali metals or other metallic elements that possess closed absorption and fluorescence cycles at a convenient wavelength.
While this technique, coupled with evaporative cooling, may lead to extremely low temperatures, revealing fascinating physical properties, the systems are of limited chemical interest.
In natural systems  metallic elements are found bonded to halogen or oxygen atoms.
Direct laser cooling of halogen atoms is difficult, however, since the lowest single-photon transition lies in the vacuum ultraviolet region of the spectrum and suitably intense laser systems are not available\cite{Kolbe2012}.

Widening the range of cold, trapped atoms and molecules will provide more than just an extension to experiments already performed with laser-cooled atoms.
Reactions of halogen atoms have been studied extensively, both theoretically and experimentally.
The low barrier to reaction with these radicals renders them suitable to the study of low-temperature kinetics.
For example, molecular hydrogen is predicted to react with fluorine with a significant rate constant at low temperatures despite the presence of an entrance-channel barrier\cite{Balakrishnan2001}.
Combining a source of cold halogens with existing cold alkali sources also offers the possibility of production of highly polar diatomic molecules, given the high electronegativity of the halogens.

Magnetic trapping of atoms is a standard route to the production of ultracold, dense, vapors.
Typically, in these experiments, electromagnetic coils generate weak trapping fields on the order of hundreds of Gauss, or superconducting\cite{Bagnato1987,Guest2005} and permanent magnets\cite{Tollett1995} provide stronger trapping fields.
Such permanent-field traps are conservative, requiring a method to remove atomic kinetic energy.
In some cases, the dissipative force can be provided by repeated absorption-emission cycles\cite{Bagnato1987}, by using a magneto-optical trap followed by pumping to a magnetically-trappable metastable state\cite{Stuhler2001}, or the trap is loaded from a buffer-gas cooled atomic\cite{Kim1997} or molecular\cite{Doyle1995} gas cell or Zeeman- or Stark-decelerated molecular beam\cite{Riedel2011,Sawyer2008}.
The difference in the present work is that the cold atoms are generated by photodissociation \emph{inside} the trap allowing, in principle, reloading and accumulation of density in a pulsed experiment, or continuous loading in a CW experiment.
It also means that the trap design can be simple (a pair of permanent magnets in this case) and compact.

Previously, we outlined an experimental scheme for production of cold halogen atoms following photodissociation of a molecular halogen precursor\cite{Doherty2011}.
Here we demonstrate for the first time the magnetic trapping of cold bromine atoms at milliKelvin temperatures, and determine the rate of trap loss due to elastic collisions with residual, room-temperature, gas in the vacuum chamber.

$\mathrm{Br}_2$ absorbs strongly above its dissociation limit in a broad continuum spanning 320--600~nm\cite{Lindeman1979}.
Two combinations of products are energetically available at wavelengths shorter than 510~nm: a pair of ground-state ($^{2}\mathrm{P}_{\frac{3}{2}}$), or one ground and one spin-orbit excited ($^{2}\mathrm{P}_{\frac{1}{2}}$) atom\cite{Cooper1998}.
The photofragment angular distribution and the contribution of each channel has been extensively studied over a wide range of wavelengths\cite{Cooper1998,Samartzis2000}, showing that dissociation into the Br~+~Br$^{*}$ channel proceeds via a parallel transition.
The resulting atoms recoil along the laser polarization axis with a near-limiting value of the anisotropy parameter, $\beta$, which characterises the photofragment angular distribution in terms of the second Legendre Polynomial $I(\theta)=\sigma/(4\pi)(1+\beta P_2(\cos\theta))$\cite{Rakitzis1999}.
Hence the laboratory-frame velocity and direction of a recoiling Br atom can be controlled via the wavelength and polarization of a photodissociation laser.

A molecular beam of $\mathrm{Br}_2$ seeded in argon expands into a vacuum chamber at 550~$\mathrm{ms}^{-1}$, and dissociation of each $\mathrm{Br}_2$ molecule at 460~nm produces a pair of atoms (Br + Br*) moving at 550~$\mathrm{ms}^{-1}$ relative to the center of mass of the parent molecule.
If the polarization of this laser pulse is parallel to the molecular beam, one of these atoms moves backwards relative to the beam direction, and cancellation between the center of mass and recoil velocities results in one atom remaining stationary in the laboratory frame of reference.
The mean velocity of the subset of backward recoiling atoms   can be arbitrarily close to zero, but in practice a distribution of finite speeds remains owing to the velocity distribution of the molecular beam and the product angular distribution of photofragmentation.
The final velocity distribution can be determined by ionizing the atoms using a (2+1) REMPI scheme and employing either ion imaging or time of flight analysis\cite{Doherty2011}.
The cloud of slow atoms will continue to expand and decrease in density, and thus a method is needed to trap the slowest atoms in a defined spatial region.

The Zeeman  splitting of the Br ground state ($^2{\rm P}_\frac{3}{2}$) is given approximately by $\Delta E = g_J m_J \mu_B |B|$, where $m_J$ can take the values $+\frac{3}{2}$, $+\frac{1}{2}$, $-\frac{1}{2}$, and $-\frac{3}{2}$ (neglecting nuclear hyperfine coupling).
Positive $m_J$ states are low-field seeking, hence atoms in these states with sufficiently low kinetic energy may be trapped near a magnetic field minimum.

\begin{figure}[htb]
    \begin{center}
        \includegraphics{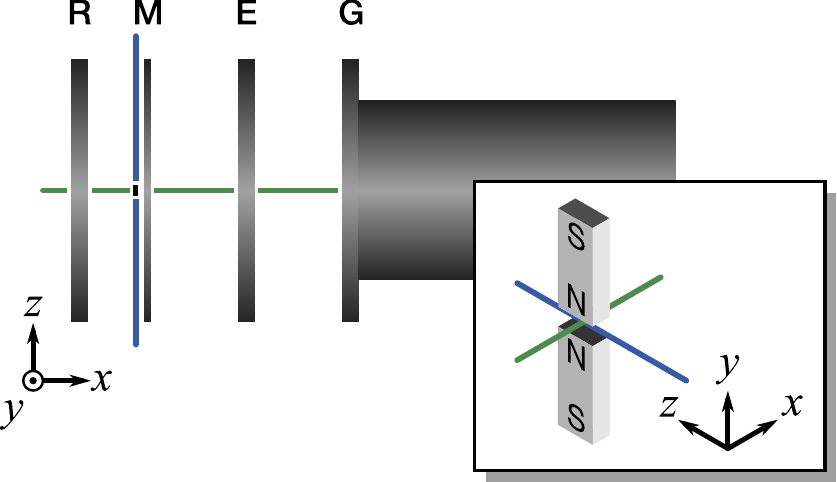}
    \end{center}
    \caption{Schematic of experimental geometry. The aluminum magnet-mount plate (M) is positioned between the repeller (R) and extractor (E) of a Wiley-McLaren electrode stack and a grounded flight-tube (G). Inset: The lasers (blue line) intersect the molecular beam (green line) between the opposing poles of a pair of permanent magnets.}
    \label{fig:expt}
\end{figure}

\begin{figure}[htb]
    \begin{center}
        \includegraphics{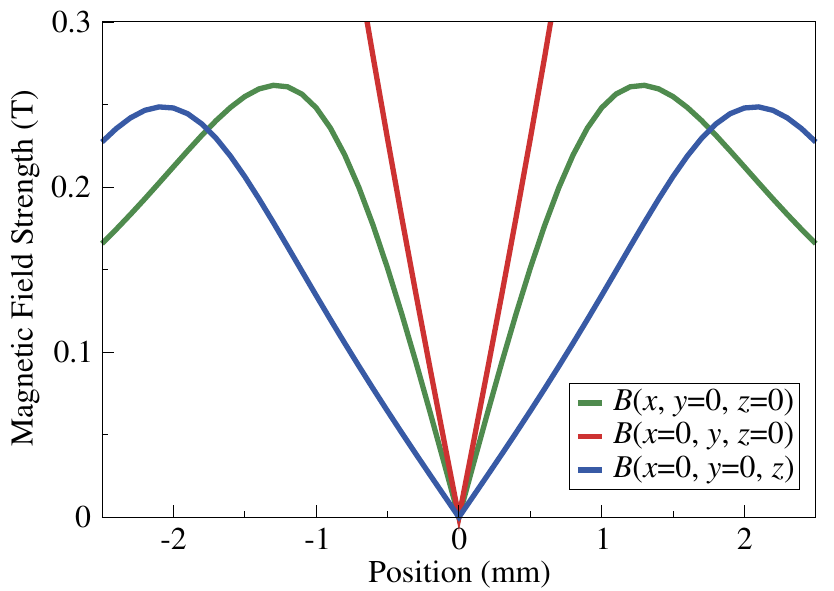}
        \caption{Magnetic field along each axis between two opposing bar magnets. Green: $x$-axis, parallel to molecular beam; red: $y$-axis, parallel to magnetization axis; blue: $z$-axis, parallel to laser propagation.}
        \label{fig:trapPotential}
    \end{center}
\end{figure}

A magnetic trap is created using a pair of opposing neodymium iron boride bar magnets (Arnold Magnetic Technologies).
The $2\times4\times10~\mathrm{mm}$ magnets are magnetized along the longest ($y$) axis with a remnance of 1.4~T, and are clamped with the north pole faces opposing and separated by 2~mm, as shown in figure \ref{fig:expt}.
The sintered N48 neodymium-iron-boron magnets have a remarkably high resistance to demagnetization -- the field due to one magnet is 0.1~T at the face of the opposing magnet, resulting in negligible demagnetization.
The components of the field can be calculated analytically\cite{Engel-Herbert2005}.
The magnetic field near the center of the trap can be approximated by adding the field components of each individual magnet, and is plotted in Figure \ref{fig:trapPotential}.
There is a field minimum  at the midpoint of the gap between the magnets, rising to a saddle point 1~mm and 2~mm from the center along the $x$ and $z$ axes respectively.

A pair of counter propagating laser beams pass through the trap center, along the $z$-axis, where they intersect  the perpendicular molecular beam, propagating along the $x$-axis (figure \ref{fig:expt}).
The first laser (a pulsed Nd:YAG-pumped dye laser) dissociates Br$_2$ seeded in an argon molecular beam, and the second (a frequency doubled dye laser) ionizes the resulting ground state atoms via a 2+1 REMPI scheme probing the  $^{2}\mathrm{P}_{\frac{3}{2}}\, \rightarrow\, ^{2}\mathrm{D}_{\frac{5}{2}}$ transition\cite{Doherty2011}.
Delaying the REMPI laser with respect to the dissociation laser probes the number of Br atoms remaining in the focal volume as a function of time.
The magnetic trap is mounted between the repeller and extractor ion optics of a standard Wiley-McLaren time-of-flight mass spectrometer that accelerates ions towards a microchannel plate detector.
The magnets are electrically biased to minimize perturbation to the local electric field and maintain extraction efficiency.

A mixture of 8\% by volume of gaseous Br$_2$ is formed by saturating argon at 4~Bar with the vapor pressure from liquid $\mathrm{Br}_2$ at room temperature.
This mixture expands through a 0.5~mm nozzle and passes through a 1~mm skimmer to the second, differentially pumped,  chamber.
This yields an estimated Br$_2$ density of $6\times10^{12}$~cm$^{-3}$ in the photodissociation focal volume.

\begin{figure}[htbp]
    \centering
    \includegraphics{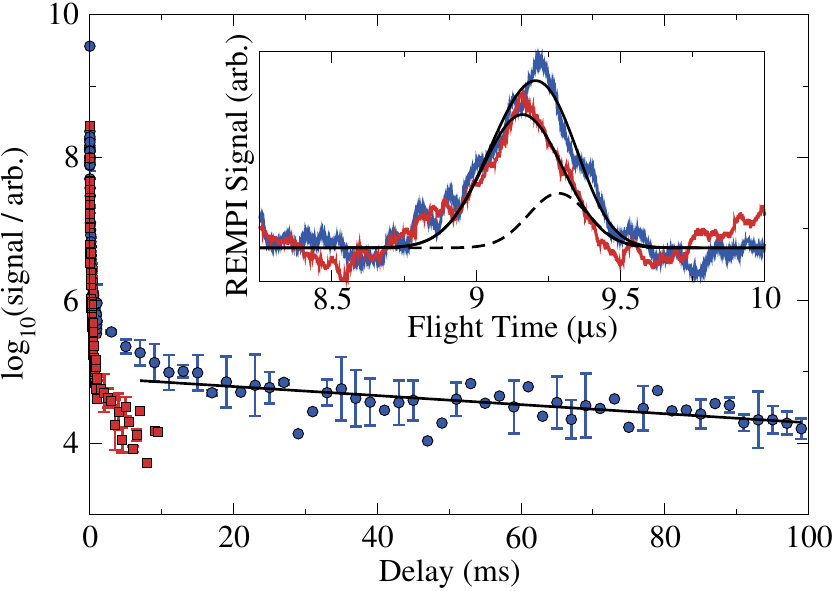}
    \caption{Integrated signal as a function of delay between dissociation and ionization, error bars are standard deviation of repeat experiments. Blue circles: magnetic field present; red squares: magnetic field absent. Linear regression on trapped signal between 5~ms and 99~ms is overlaid in black. Inset: representative time-of-flight trace recorded at a photodissociation to REMPI laser delay of 51~ms. Red trace: photodissociation laser blocked; blue trace: photodissociation laser unblocked. The black lines are Gaussian fits, and the dashed black line is the signal attributable to trapped bromine atoms.}
    \label{fig:fullDecay}
\end{figure}

The probe laser wavelength is sufficiently short to dissociate residual $\mathrm{Br}_2$ and then ionize the resulting fragments.
Even at the longest delays probed here this one-color signal remains a significant contribution to the background and dominates the overall signal.
While some of this contribution can be eliminated based on the ion flight time -- these atoms have much higher initial velocity than the near-zero speed of trapped atoms -- the angular anisotropy of the unwanted photofragments still produces atoms with a zero velocity component in the time-of-flight direction.
This contribution to the detected signal is eliminated by recording two alternating batches of TOF traces: a 10-shot average of the two-color TOF trace, followed by a 10-shot average recorded using only the probe laser. These signals are shown inset in figure \ref{fig:fullDecay} as the blue and red traces respectively.

The difference between these two traces is attributed to Br atoms produced at near-zero velocity that have remained in the probe region until ionized and detected.
To extract a signal proportional to the number of atoms remaining, we adopt a maximum likelihood fitting technique that accounts for the statistics of ion counting, and effectively recovers a small signal arriving on top of a large background\cite{Hannam1999}.

An extra signal is clearly visible above the one-color background in the trace inset in figure \ref{fig:fullDecay}.
Integrating this peak gives an estimate of the number of Br atoms remaining in the probe volume after a given time delay.
Figure \ref{fig:fullDecay} plots (in blue circles) this integrated signal as a function of the delay between the dissociation and probe laser pulses.
The confidence interval of the integrated signal from a single experiment is smaller than the marker in the figure, and the error bars indicate the standard deviation of repeated experiments.

The integrated signal shows a rapid loss of atoms over the first few hundred nanoseconds, transitioning to a slower, exponential loss after this.
The rapid early decay of signal  reflects the unhindered ballistic expansion of the fastest atoms produced on photodissociation.
As the cloud of photodissociation products expands, successively slower atoms leave the probe volume and are irreversibly lost as their kinetic energy exceeds the trap depth.
The inhomogeneous magnetic field will trap faster atoms until their trajectory takes them through the saddle point along the $z$ axis.
After 1 ms, the remaining atoms are those with sufficiently slow initial velocities to be trapped in the magnetic field and the exponentially-decaying signal is indicative of a first-order process.

Replacing the neodymium ceramic magnets with aluminum pieces removes the magnetic field from the interaction region while maintaining the local electric field.
The relative density of these atoms as a function of delay time is plotted as red squares in figure \ref{fig:fullDecay}.
The notable difference between the two signals is that no atoms were detected beyond 5~ms with no magnetic field present; there is no statistically-significant additional signal when the dissociation laser was fired.
This measurement demonstrates that the magnetic trap is responsible for the long tail in the signal out to 100 ms.

The significance of the observation of signal out to 100~ms is that with a laser repetition rate of 10~Hz, there will be trapped population from the previous pulse when a subsequent photolysis pulse reloads the trap.
Unlike many other methods, reloading can be performed without ``opening'' the trap as the cold atoms are formed within the trap.

Collisions of the trapped atoms with the tail of the molecular beam contribute to the earliest losses, but the primary loss mechanism is via inelastic collisions with residual gas in the vacuum chamber.
The chamber pressure rises to $4\times10^{-7}$~mbar during the experiment, reaching an equilibrium between the gas added via the molecular beam (primarily the argon seed gas) and the pumping speed of the turbomolecular pump.
This background presents an average density of $9.7\times10^{9}$~cm$^{-3}$ with a room-temperature Maxwell-Boltzmann velocity distribution.

A Br atom will leave the trap if the collision has transferred sufficient kinetic energy to overcome the magnetostatic potential.
The energy transferred depends on the geometry of the collision as $\Delta E \approx \frac{\mu^{2}}{m_2} |v_r|^{2}(1-\cos \theta)$ \cite{Fagnan2009}, where $\theta$ is the collision angle between the initial and final velocity vectors.
Equating the trap depth to $\Delta E$ defines a minimum collision angle, $\theta_{\mathrm{min}}$, which depends on the relative collision velocity.
The total cross section for trap loss is given by the integral of the differential cross section (DCS) using the minimum collision angle as a lower bound:
\begin{equation}
\sigma(v_{\mathrm{Ar}}) = \int_{\theta_{\mathrm{min}}(v_{\mathrm{Ar}})}^{\pi} 
(d\sigma/d\Omega)\,d\Omega.
\label{equ:dcs}
\end{equation}
The trap ejection rate constant is then given by the velocity-averaged collision cross section, $k' = \langle v\sigma\rangle$:
\begin{multline}
    \langle v\sigma \rangle = 4\pi\left( \frac{m_{\mathrm{Ar}}}{2\pi 
        k_{\mathrm{B}} T} \right)^{\frac{3}{2}} \\ 
        \int_0^{\infty} \sigma(v_{\mathrm{Ar}})\, v_{\mathrm{Ar}}^{3}\, 
        \exp\left\{ -\frac{m_{\mathrm{Ar}} v_{\mathrm{Ar}}^{2}}{2 k_{\mathrm{B}} 
    T} \right\}\,d v_{\mathrm{Ar}},
    \label{equ:vsigma}
\end{multline}
integrating over a Maxwell-Boltzmann velocity distribution at room temperature.

Collisions between rare gas and halogen atoms have been well studied at these energies yielding experimentally-determined potential energy curves\cite{Casavecchia1981}.
In this work, to calculate the DCS, elastic scattering wavefunctions are propagated on these radial potential energy curves using the log-derivative method\cite{Alexander1987}.
Matching the log-derivative matrix to the asymptotic solutions at large interatomic separation yields the scattering matrix, from which the DCS is extracted.

Trap loss kinetics can be approximated by a pseudo-first order mechanism in which the density of background argon is considered to be constant with time: $d[\mathrm{Br}]/dt = k [\mathrm{Br}]$.
The loss rate is then characterized by a rate constant given by the product of the argon density and a collision cross section, $k =\langle v\sigma \rangle[\mathrm{Ar}]$.
Ignoring the first 5 milliseconds of data due to competing loss processes, the decay from 5~ms to 100~ms is well-described by a single exponential and fits to a decay rate of $(12.7\pm2.6)\, \mathrm{s}^{-1}$.
Including an estimated 10\% error in the chamber pressure, the measured velocity-averaged cross section is $\langle v\sigma\rangle = (1.31\pm0.29)\times10^{-9}\,\mathrm{cm}^{-3}\mathrm{s}^{-1}$.
Using this value in equations (\ref{equ:vsigma}) and (\ref{equ:dcs}) defines a value for $\theta_{\mathrm{min}}$, and hence also the trap depth, $U_0 / k_{\mathrm{B}} = (270\pm35)\,\mathrm{mK}$.

The trap depth corresponds to the potential energy of an atom at the lowest saddle-point in the trapping potential, seen in figure \ref{fig:trapPotential} to correspond to a field strength of $(0.24\pm0.02)\,\mathrm{T}$, determined from the manufacturer's nominal magnetization and tolerance.
The potential energy of a Br atom in the $m_j=\frac{3}{2}$ sub-state at this location is $U/k_\mathrm{B} = (322\pm36)\,\mathrm{K}$, in good agreement with our measured trap depth.
The spin-orbit excited state possess a Land\'e g-factor half the magnitude of the ground state, resulting in a much shallower trap depth of $U/k_\mathrm{B} = (56\pm4)\,\mathrm{K}$ for the $m_j=\frac{1}{2}$ sub-state.

Photodissociation of Br$_2$ produces an anisotropic velocity distribution containing speeds in the lab frame ranging from 0~ms$^{-1}$ to 1120~ms$^{-1}$.
No additional cooling mechanism operates after photodissociation, and the trap only retains the fraction with a low enough initial speed.
The estimated trap depth of 270~mK corresponds to a maximum speed of 7.5~$\mathrm{ms}^{-1}$ for a Br atom.
Integrating the distribution of photofragments formed below this limit leads to the conclusion that $4.1\times10^{-3}$ of the photodissociation yield is trappable, which is consistent with the initial decrease in signal seen in figure \ref{fig:fullDecay}.
The dissociation laser intensity is sufficient to saturate the transition which produces ground state Br atoms at a 60\% branching fraction.
Of these atoms we assume a statistical distribution between the $m_j$ sub-levels.
The 5~ns, pulsed, photodissociation laser is focused into the magnetic field minimum, effectively illuminating a 750~$\mu$m diameter cylinder though the center of the trap.
The peak density of atoms formed here expands to fill the trap on a microsecond timescale, reducing the density of Br atoms by a factor of 2.6.
Integrating the trapped fraction over the illumination volume of the 
dissociation laser, we estimate an upper limit on trapped ground-state atom 
density of $1.3\times 10^{8}$~cm$^{-3}$ and $3.6\times 10^{8}$~cm$^{-3}$ in the 
$m_j=\frac{1}{2}$ and $m_j=\frac{3}{2}$ sub-states respectively, a total of 
$2\times10^{6}$ gound-state atoms.
Under these conditions a density of $6.4\times 10^{7}$~cm$^{-3}$ spin-orbit excited $^{2}\mathrm{P}_{\frac{1}{2}}$ atoms is also trapped.
The detection laser is tuned to resonantly ionize only the ground $^{2}\mathrm{P}_{\frac{3}{2}}$ state, so the Br$^*$ atoms do not contribute to the measured signal.
They have been detected in the trap, but the density is found to be much lower than Br.

Trapped $^{2}\mathrm{P}_{\frac{1}{2}}$ atoms are metastable with respect to radiative decay and collisional quenching\cite{Johnson1996}.
The radiative lifetime is 1.12~s, and the trapped atom density is sufficiently low that spin-orbit exchange collisions between Br and Br$^*$ are unlikely.
We do not expect to see any effects arising from repopulation of the ground state on the 100~ms timescale of these experiments.

While it would be feasible in principle to load the trap continuously rather than in a pulsed format -- dissociating the parent Br$_2$ molecule from a continuous beam using a sufficiently powerful diode or CW dye laser -- the loading rate would be limited by collisions between the high-density molecular beam and trapped atoms, and the background pressure would be much higher unless very large pumping systems were employed.
In the pulsed case we can time the dissociation near the tail of the molecular beam pulse, minimizing such losses.
Detection of trapped Br atoms is also rendered relatively straightforward by multiphoton ionization, which requires high-power, and hence pulsed, ultraviolet light.
Shortening the molecular beam pulse will enable a reduction in the collisions that contribute to trap loss and thus a genuine accumulation of signal over multiple pulses.
Increasing the phase space density by sympathetic cooling with ultracold H atoms may be a possibility, as suggested recently by calculations for F atoms\cite{Hutson2013}.

In summary, we have demonstrated magnetic trapping of Br atoms for the first time, and these are held for sufficiently long (100~ms) to enable reloading of the trap in a photodissociation experiment operating at 10~Hz.
In principle this experiment can be extended to trapping other halogen atoms such as Cl by photolysis of Cl$_2$ or F by photolysis of ClF, or N or O via photolysis of NO or NO$_2$.
The only limitation is finding a precursor that absorbs at an appropriate energy near a photodissociation limit to yield fragments with required velocities to cancel the parent velocity.
Accumulation over repeated loading cycles following dissociation by high-power pulsed lasers will allow the use of parent molecules with low absorption coefficients, or with low branching ratios into the required channels.

\bibliography{ref}
\end{document}